\documentclass[3p,times]{elsarticle}

\usepackage{ecrc}

\volume{00}

\firstpage{1}

\journalname{Nuclear Physics A}

\runauth{D. Jido et al.}

\jid{NUPHA}

\jnltitlelogo{Nuclear Physics A}

\CopyrightLine{2012}{Published by Elsevier Ltd.}

\usepackage{amssymb}

\usepackage[figuresright]{rotating}

\begin{document}

\begin{frontmatter}

\dochead{}



\title{$\eta^{\prime}$ meson under partial restoration of chiral symmetry in nuclear medium}

\author[YITP]{Daisuke Jido}
\author[Kyoto]{Shuntaro Sakai}
\author[Nara]{Hideko Nagahiro}
\author[Nara]{Satoru Hirenzaki}
\author[Nara]{Natsumi Ikeno}

\address[YITP]{Yukawa Institute for Theoretical Physics, Kyoto University, Kyoto 606-8502, Japan}
\address[Kyoto]{Department of Physics, Graduate School of Science, Kyoto University, Kyoto, 606-8502, Japan}
\address[Nara]{Department of Physics, Nara Women's University, 
Nara 630-8506, Japan}

\begin{abstract}
In-medium modification of the $\eta^{\prime}$ mass is discussed in the context of partial restoration of chiral symmetry in nuclear medium. We emphasize that the U$_{A}$(1) anomaly effects causes the $\eta^{\prime}$-$\eta$ mass difference necessarily through the chiral symmetry breaking. As a consequence, the $\eta^{\prime}$ mass is expected to be reduced by order of 100 MeV in nuclear matter where about 30\% reduction of chiral symmetry takes place. The strong attraction relating to the $\eta^{\prime}$ mass generation eventually implies that there should be also a strong attractive interaction in the scalar channel of the $\eta^{\prime}$-$N$ two-body system. We find that the attraction can be strong enough to form a bound state. 
\end{abstract}


\end{frontmatter}


\section{Introduction}
\label{intro}
Recently the properties of the $\eta^{\prime}$ meson have been studied 
intensively in both theoretical and experimental aspects. Particularly 
its in-medium properties are very interesting in the context of exploring 
possible bound states of the $\eta^{\prime}$ meson in nuclei as suggested 
in Refs.~\cite{Nagahiro:2004qz,Nagahiro:2006dr}. It is known that 
the flavor singlet pseudoscalar meson strongly depends on the breaking pattern
of chiral symmetry in QCD~\cite{Cohen:1996ng,Lee:1996zy}. Especially
in the SU(3) chiral symmetry the flavor octet and singlet pseudoscalar mesons 
belong to the same SU(3) chiral multiplet and they should degenerate 
each other in the chiral symmetric limit together with the scalar partners
even though the U$_{A}$(1) symmetry is broken. This feature 
implies that the $\eta^{\prime}$ mass is expected to be reduced
under partial restoration of chiral symmetry in nuclear medium~\cite{Jido:2011pq}.
In Ref.~\cite{Oset:2010ub} $\eta^{\prime}$-nucleon scattering is discussed
based on a coupled channel approach with chiral dynamics. Using this formulation
Ref.~\cite{Nagahiro:2011fi} has calculated the $\eta^{\prime}$ optical 
potential in nuclei for one and two-body absorptions. 
Experimentally the observation of the transparency ratio 
in the $\eta^{\prime}$ photoproduction on nuclear targets by 
the CBELSA/TAPS collaboration~\cite{Nanova:2012vw} 
is so remarkable that the observed transparency 
is sufficiently larger than those of the $\omega$ and $\eta$ mesons. 
This result estimates the in-medium $\eta^{\prime}$ width to be 15-25 MeV 
at the saturation density for an average momentum 1 GeV/c independently of
momenta in a certain range. Thus, the absorption of the $\eta^{\prime}$ meson
in nuclear matter may be small. 
The elastic $\eta^{\prime}N$ interaction could
be also weak according to the threshold $\eta^{\prime}$ production from
$pp$~\cite{Moskal:2000gj}.
Nevertheless, a recent theoretical calculation based on chiral dynamics with many-body treatment up to the second order has suggested
that both observations could be hard to be reproduced simultaneously~\cite{Nagahiro:2011fi}.
An experimental observation of $\eta^{\prime}$-nucleus bound states 
by a $(p,d)$ reaction on a $^{12}$C target is planned at GSI~\cite{Itahashi:2012ut}.
The formation spectrum of the $(p,d)$ reaction is comprehensively investigated with 
various attraction strengths of the $\eta^{\prime}$ meson in nucleus 
in Ref.~\cite{Nagahiro:2012}.


\section{$\eta^{\prime}$ mass in nuclear medium}
\begin{figure}[t]
\begin{center}
\includegraphics[bb=0 0 875 524, width=0.55\columnwidth]{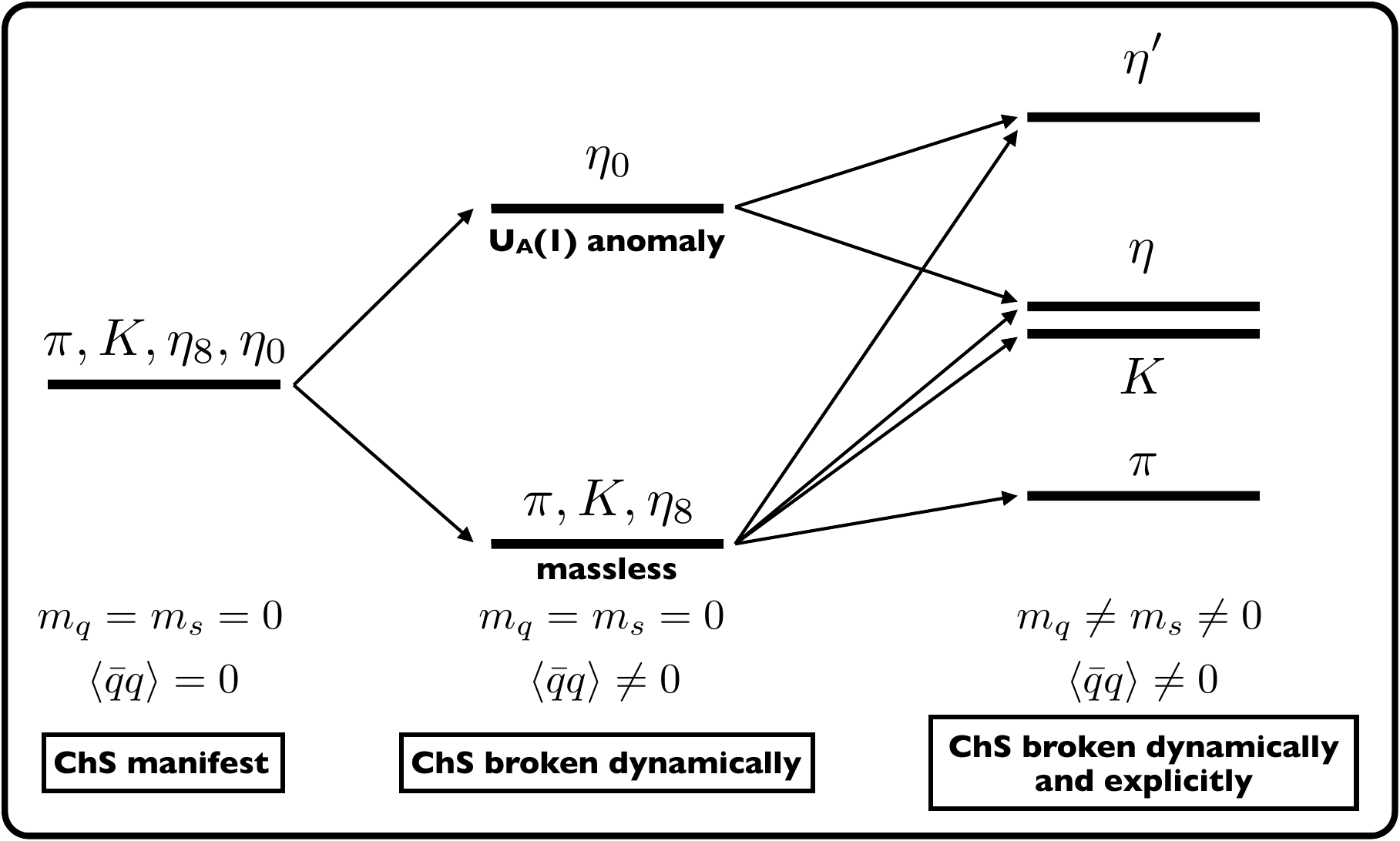}
\includegraphics[bb=0 0 586 463, width=0.43\columnwidth]{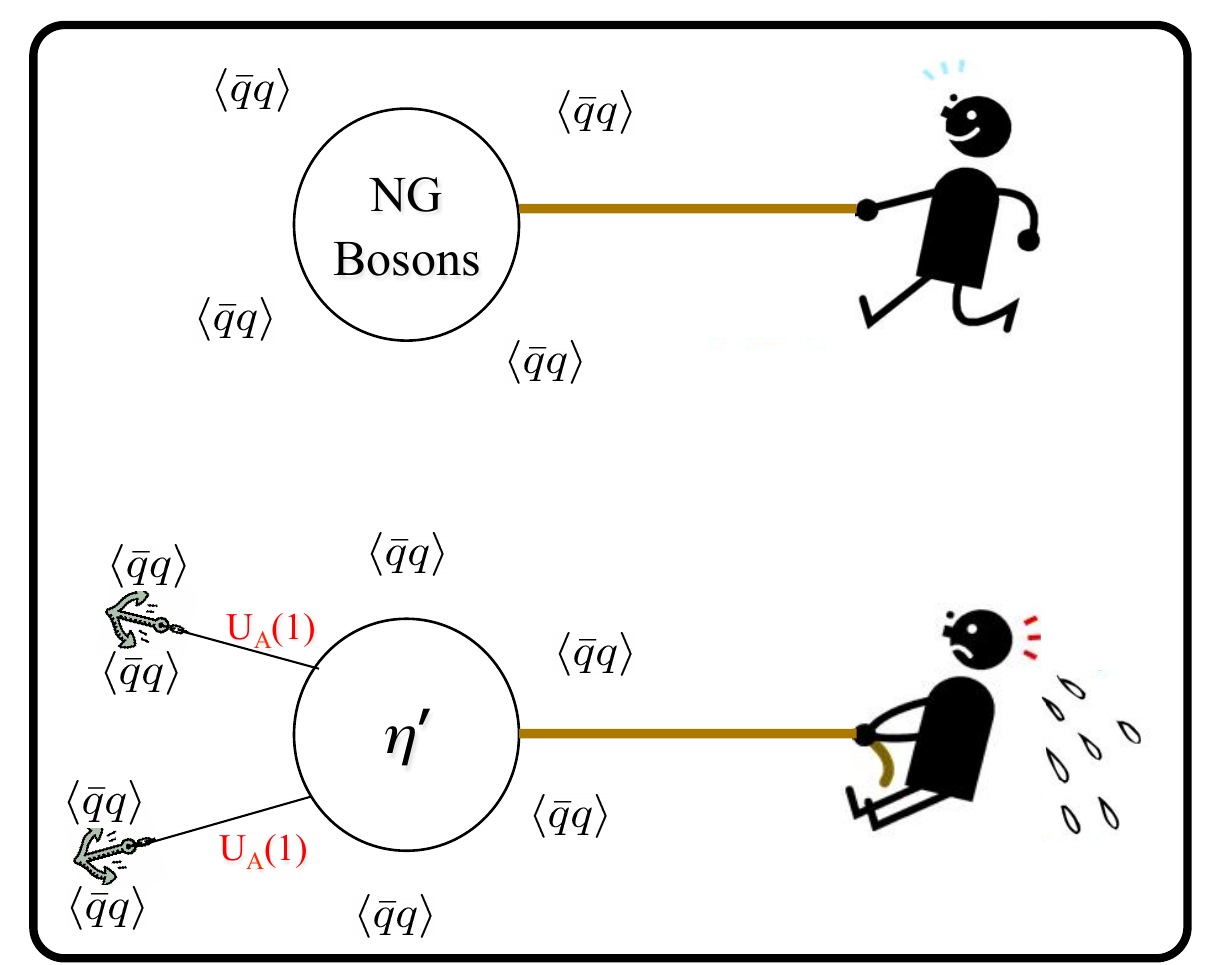}
\end{center}
\caption{(Left) Light pseudoscalar meson spectrum in the various patterns of the SU(3) chiral symmetry breaking. In the left,  chiral symmetry is manifest without explicit nor dynamical breaking. All the pseudoscalar mesons have a common mass. In the middle, chiral symmetry is dynamically broken in the chiral limit.
The octet pseudoscalar mesons are identified as the Nambu-Goldstone bosons associated with the symmetry breaking.  In the right, chiral symmetry is broken dynamically and explicitly.
The U$_{A}$(1) symmetry is always broken without assuming its breaking pattern. 
(Right) Intuitive picture of the role of the U$_{A}$(1) anomaly and the chiral symmetry breaking for the Nambu-Goldstone bosons and the $\eta^{\prime}$ meson. The mass of the Nambu-Goldstone boson is light owing to the spontaneous symmetry breaking, while the $\eta^{\prime}$ meson is anchored by the U$_{A}$(1) anomaly to the vacuum condensate. }
\label{fig:1}       
\end{figure}

It is the well-known fact that the U$_{A}$(1) symmetry in QCD is broken 
explicitly by the quantum anomaly. Consequently the flavor singlet 
pseudoscalar meson $\eta_{0}$ is not a Nambu-Goldstone boson associated 
with the spontaneous breaking of chiral symmetry any more.
Eventually the $\eta^{\prime}$ meson 
has a larger mass than the other pseudoscalar mesons 
being the Nambu-Goldstone bosons. 
In fact the $\pi$, $\eta$ and $\eta^{\prime}$ 
with the strangeness partners belong to the same multiplet 
of the chiral SU(3) group together with the scalar companies,
thus in the SU(3) chiral symmetric limit these particles get 
degenerated~\cite{Lee:1996zy,Jido:2011pq},  as shown in Fig.~\ref{fig:1} (left),
even though the U$_{A}$(1) symmetry is broken owing to the quantum anomaly. 
This implies that 
the mass gap between $\eta$ and $\eta^{\prime}$ is generated necessarily by 
the SU(3) chiral symmetry breaking in the sense of both explicitly and
spontaneously under the presence of the U$_{A}$(1) breaking as depicted 
in Fig.~\ref{fig:1} (right). 
Once assuming that the spontaneous 
SU(3) chiral symmetry breaking predominates over the explicit breaking by
the strange quark, one expects that the $\eta^{\prime}$ and $\eta$
mass difference should decrease as the chiral symmetry is being restored. 
Therefore under partial restoration of chiral symmetry in nuclear matter
the $\eta^{\prime}$ meson mass gets reduced since the $\eta$ mass
hardly changes in nuclear matter owing to its Nambu-Goldstone
boson nature. The magnitude of the mass reduction is estimated as an order of 100 to 150
MeV at the saturation density~\cite{Jido:2011pq}. 

The fact that the flavor singlet pseudoscalar $\eta_{0}$ belongs to 
the same multiplet of the SU(3) chiral group as the octet pseudoscalars
can be shown in the following argument:
The pseudoscalar and scalar meson fields are composed by 
pairs of $\bar q_{L}q_{R}$ and $\bar q_{R}q_{L}$  as 
$ \phi, \phi_{5} = \bar q_{L}q_{R}\pm\bar q_{R} q_{L}$. 
Since the left and right quarks are in the fundamental representation of SU(3), $\bf 3$, 
the combination $\bar q_{L}q_{R}\oplus \bar q_{R} q_{L}$ belongs to the $({\bf \bar 3}, {\bf 3}) \oplus
({\bf 3}, {\bf \bar 3})$ multiplet. The member of this multiplet can be classified 
into the terminology of  SU(3)$_{V}$ which is a subgroup of SU(3)$_{L}\otimes$SU(3)$_{R}$. Since the vector transformation does not distinguish the left and right
rotations, we have the singlet and octet representations of SU(3)$_{V}$ in the
$({\bf \bar 3}, {\bf 3}) \oplus ({\bf 3}, {\bf \bar 3})$ multiplet as obtained by
${\bf \bar 3} \otimes {\bf 3} = {\bf 1} \oplus {\bf 8}$.
Therefore, we have $\eta_{8}$ and $\eta_{0}$ in the same multiplet. 
We also show that these fields can be transformed each other under the
axial transformations: Since the axial transformation is an adjoint (octet) 
operator of SU(3)$_{V}$ and is not a generator of SU(3)$_{V}$, 
it can mix the octet and singlet, for instance 
$[Q_{A}^{a}, \eta_{0}]=\sqrt{2/3} \phi^{a}$ meaning 
that the singlet pseudoscalar is transformed into an octet
scalar under the axial transformation $Q_{A}^{a}$. The scalar octet 
is also transformed into a linear combination of the singlet and octet pseudoscalars 
as $[Q_{A}^{a}, \phi^{b}]=\delta^{ab}\sqrt{2/3}\eta_{0}+d^{abc} \phi_{5}^{c}$
with the SU(3) structure constant $d^{abc}$. 
In this way, the singlet and octet pseudoscalar 
fields are transformed each other under twice axial transformations, and thus
are in the same multiplet even though the U$_{A}$(1) symmetry is broken.


\section{Linear $\sigma$ model and $\eta^{\prime}$-$N$ interaction}
Let us consider the SU(3) linear $\sigma$ 
model~\cite{Lenaghan:2000}
as a chiral effective model embodying partial restoration of chiral symmetry.
\begin{eqnarray}
\mathcal{L}&=&{\textstyle \frac{1}{2}}\mathrm{tr}\partial_\mu M\partial^\mu M^\dagger
 -{\textstyle \frac{{\mu}^2}{2}} \mathrm{tr}(MM^{\dagger})
 -{\textstyle \frac{\lambda}{4}}\mathrm{tr}[(MM^{\dagger})^2] 
 -{\textstyle \frac{{\lambda}^{\prime}}{4}} [\mathrm{tr}(MM^{\dagger})]^2 \nonumber\\
& &-A\,\mathrm{tr}( {\chi} M^{\dagger} + \chi^\dagger M ) 
+{\textstyle \sqrt{3}}B( \det M + \det M^\dagger ) \, \label{LsigLag}
\end{eqnarray}
where the meson field is given by $M=\sum^8_{a=0} (\lambda_a \sigma_a +i\lambda_a \pi_a)/\sqrt{2} $ with the scalar and pseudoscalar mesons, $\sigma_{a}$ and $\pi_{a}$, respectively, and the Gell-Mann matrix $\lambda_{a}$. $M$ and $\chi$ belong to the $({\bf \bar 3}, {\bf 3})$ multiplet transforming $(M,\chi) \to L (M,\chi) R^{\dagger}$ under the chiral rotation. By fixing $\chi = \mathrm{diag}(m, m, m_s)$ with the $u$-$d$ and strange quark masses, $m$ and $m_{s}$, the chiral and flavor symmetries are explicitly broken. The last term with the parameter $B$ of Eq.~(\ref{LsigLag}) represents the axial anomaly effect breaking the U$_A$(1) symmetry.
For certain parameter sets, chiral symmetry is spontaneously broken with a finite chiral condensate given by a linear combination of $\langle \sigma_{0}\rangle$ and $\langle \sigma_{8}\rangle$, and the octet pseudoscalar mesons become the Nambu-Goldstone bosons. The $\eta^{\prime}$ mass in the chiral limit is obtained as 
\begin{equation}
m_{\eta^{\prime}}^{2}=6B\langle \sigma_{0} \rangle
\end{equation}
at tree level. This implies that the $\eta^{\prime}$ mass is generated by the chiral symmetry breaking as well as the anomaly effect. Consequently, when chiral symmetry is partially restored in nuclear matter, the $\eta^{\prime}$ mass should be reduced.

The nuclear medium effect is introduced based on the SU(3) linear sigma model with the baryon field~\cite{Kawarabayashi:1980uh,Sakai}. An effective Lagrangian for the meson in linear density~\cite{Sakai} is obtained as
\begin{equation}
   \mathcal{L}_{\rm MF} =  - \frac{g\rho}{\sqrt{3}} 
   \left(\sigma_0 +\frac{\sigma_8}{\sqrt{2}}\right)
    - \frac{1}{2} \frac{g^{2}\rho}{m_{N}}
     \left( \frac{1}{6} \eta_{8}^{2} + \frac{1}{3} \eta_{0}^{2}
    + \sqrt 2 \frac{2}{3} \eta_{0} \eta_{8}\right) \ , \label{MF}
\end{equation}
with the nuclear density $\rho$  and  the coupling constant $g$ for the nucleon meson Yukawa interaction. The second term of Eq.~(\ref{MF}) comes from the particle-hole excitation in the meson propagation. We have omitted other interactions  irrelevant to the current discussion. The vacuum is changed owing to the first term at finite density. The in-medium modification of the meson masses stems from two contributions, $m^{*2}=m^{2}(\langle \sigma_{0}\rangle^{*}, \langle \sigma_{8}\rangle^{*}) + \Sigma_{\rm ph}$: one is the effect of the shift of the vacuum induced by the first term of Eq.~(\ref{MF}), and the other is the particle-hole excitations expressed by the second term of Eq.~(\ref{MF}). Determining the value of $g$ such that  partial restoration of chiral symmetry takes place with 35\% reduction of the quark condensate at the saturation density, we find the reduction of the $\eta$-$\eta^{\prime}$ mass difference to be 135~MeV at the saturation density~\cite{Sakai}. 


\begin{figure}[t]
\begin{center}
\includegraphics[bb=0 0 776 138, width=0.76\columnwidth]{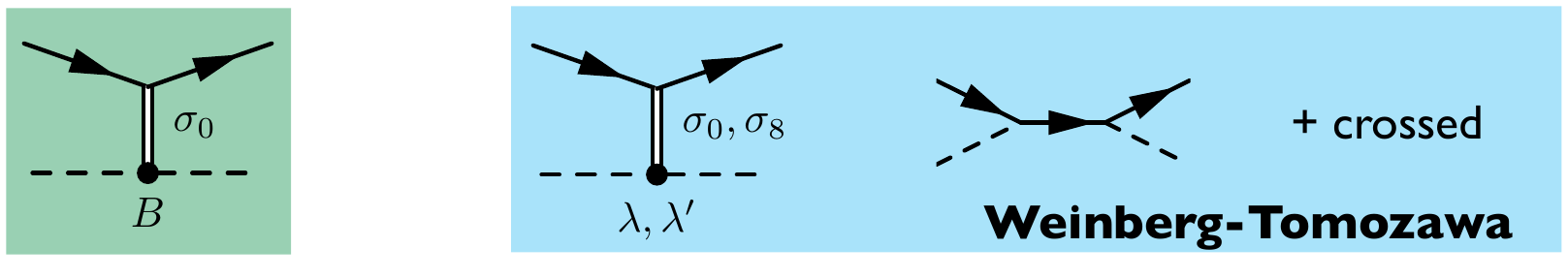}
\end{center}
\caption{Diagrammatic representation of the $\eta^{\prime}$-nucleon interaction. The left diagram comes from the $\sigma$ meson exchange with the anomalous $\eta_{0}\eta_{0}\sigma_{0}$ coupling. The residual terms cancel in low energy, giving the  Weinberg-Tomozawa interaction, which is null for the $\eta^{\prime}N$ channel. }
\label{fig:2}       
\end{figure}

The $\eta^{\prime}$-$N$ two-body interaction can be extracted from the above linear $\sigma$ model~\cite{Sakai}. At the tree level, there are two types of the contributions, the $\sigma$ meson exchange and the Born term, as depicted in Fig.~\ref{fig:2}. It is notable that, thanks to chiral symmetry, the sum of a part of the $\sigma$ meson exchange and the Born terms give the Weinberg-Tomozawa interaction in low energy and it is known to be null for the $\eta^{\prime}N$ channel from the flavor structure. Therefore, the contributions coming from the right three terms of Fig.~\ref{fig:2} cancel at low energy in the chiral limit and there remains the $\sigma$ exchange term with the $\eta_{0}\eta_{0}\sigma_{0}$ coupling coming from the anomaly effect as shown in the left diagram of Fig.~\ref{fig:2}. The $\eta^{\prime}$-$N$ interaction from the $\sigma$ exchange is estimated to be $V_{\eta^{\prime}N}=-6B  (g/\sqrt 3)/m_{\sigma}^{2}$, which is $-0.053$~MeV$^{-1}$ in the parameter producing 35\% reduction of the quark condensate at the saturation density. This interaction is comparably strong to the $\bar KN$ interaction with $I=0$ obtained by the Weinberg-Tomozawa term as $ V_{\bar KN}=-0.086$ MeV$^{-1}$ at the threshold, which dynamically generates the $\Lambda(1405)$ as a quasibound state with a 10-15 MeV binding energy. Using the same machinery as in the $\Lambda(1405)$ in the $\bar KN$ channel,  we calculate 
the $s$ wave amplitude
with the $\eta^{\prime}N$ interaction obtained here, and find that $\eta^{\prime}$ and $N$ form a bound state with a 6 MeV binding energy, and that the scattering length is order of 2 fm with the repulsive sign, which is not consistent with the value extracted from the $\eta^{\prime}$ production in the $pp$ collision~\cite{Moskal:2000gj}. Here we have used the natural renormalization scheme suggested in Ref.~\cite{Hyodo:2008} in order to exclude other components than $\eta^{\prime}$ and $N$. In this way the $\eta^{\prime}N$ system has strong attraction in the scalar channel.

It is worth emphasizing again that the mass gap between $\eta$ and $\eta^{\prime}$ stems from the SU(3) chiral symmetry breaking, especially the spontaneous breaking owing to the modest explicit breaking by the strange quark mass. Thus, a good part of the $\eta^{\prime}$ mass is generated by the sigma condensate in the view of the linear sigma model. This leads to a large coupling of the $\eta^{\prime}\eta^{\prime}\sigma$ vertex as one of the U$_{A}$(1) anomaly effects. This is the reason that the scalar channel of the $\eta^{\prime}N$ interaction is strongly attractive. This scenario recalls us of the $NN$ interaction in the scalar-isoscalar channel. In the linear sigma model, the nucleon mass is explained as a consequence of spontaneous breaking of chiral symmetry with a finite sigma condensate and the nucleon has a strong Yukawa coupling with the sigma meson. Consequently the $NN$ interaction has a strong attraction in the scalar-isoscalar channel induced by the sigma meson exchange. Therefore, the mechanism of the strong attraction of the $\eta^{\prime}N$ system in the scalar channel is same as the $NN$ attraction, and it would be natural to expect a bound state of $\eta^{\prime}$ and $N$ in a similar way to the deuteron as a $pn$ bound state. For the detailed discussion, one should examine the $\eta^{\prime}N$ interactions in the other channels. 

\section{Conclusion}
We have discussed the $\eta^{\prime}$ meson in the context of partial restoration 
of chiral symmetry in the nuclear medium. Emphasizing the significant role of 
the chiral symmetry (spontaneous) breaking for the generation of the 
$\eta^{\prime}$-$\eta$ mass gap, we find
a substantial reduction of the $\eta^{\prime}$-$\eta$ mass difference 
in the nuclear medium with an order of 100-150 MeV at the 
saturation density. We also find that, 
since the mass generation of the $\eta^{\prime}$ mass
is a consequence of the spontaneous chiral symmetry breaking with a help of 
U$_{A}$(1) anomaly,
$\eta^{\prime}$ has a strong coupling to the $\sigma$ meson. This suggests
that there should be a large attraction between $\eta^{\prime}$ and $N$
in the scalar channel with the $\sigma$ meson exchange as
the same mechanism to the attraction of the $NN$ interaction 
in the isoscalar-scalar channel.
Thus, if $\eta^{\prime}$-$N$ and $\eta^{\prime}$-nucleus systems have
weak interactions, 
there should be repulsive interactions in other channels than 
the scalar channel. Nevertheless, there is no vector interaction for the
$\eta^{\prime}N$ channel induced by the Weinberg-Tomozawa interaction, 
since it is not allowed by the flavor structure of the $\eta^{\prime} N$ system. 
For the detailed discussion, it is certainly necessary to proceed more 
phenomenological studies, such as reproduction of 
the $\eta^{\prime}$ production reactions using chiral perturbation 
theory implementing the aspect discussed here. 

\section*{Acknowledgments}
This work was partially supported by the Grants-in-Aid for Scientific Research (No.\ 22740161 and No.\ 24540274).
A part of this work is done under Yukawa International Project for 
Quark-Hadron Sciences (YIPQS).









%
%

\end{document}